\documentclass[twocolumn,epjc3]{svjour3}
\smartqed  
\usepackage{amssymb,amsmath}
\usepackage{graphicx}
\usepackage{hyperref}

\newcommand\rf[1]{(\ref{eq:#1})}
\newcommand\lab[1]{\label{eq:#1}}
\newcommand\nonu{\nonumber}
\newcommand\br{\begin{eqnarray}}
\newcommand\er{\end{eqnarray}}
\newcommand\be{\begin{equation}}
\newcommand\ee{\end{equation}}

\newcommand\foot[1]{\footnotemark\footnotetext{#1}}
\newcommand\lb{\lbrack}
\newcommand\rb{\rbrack}
\newcommand\llangle{\left\langle}
\newcommand\rrangle{\right\rangle}

\renewcommand\({\left(}
\renewcommand\){\right)}
\newcommand\bv{\bigm\vert}               
\newcommand\bgv{\bigg\vert}              

\newcommand\bc{\begin{center}}
\newcommand\ec{\end{center}}

\newcommand\partder[2]{\frac{{\partial {#1}}}{{\partial {#2}}}}

\renewcommand\a{\alpha}

\renewcommand\d{\delta}

\newcommand\eps{\epsilon}
\newcommand\vareps{\varepsilon}

\newcommand\G{\Gamma}

\newcommand\h{\frac{1}{2}}
\renewcommand\k{\kappa}
\renewcommand\l{\lambda}
\renewcommand\L{\Lambda}
\newcommand\m{\mu}
\newcommand\n{\nu}
\newcommand\om{\omega}

\newcommand\vp{\varphi}
\renewcommand\P{\Phi}
\newcommand\pa{\partial}

\newcommand\pr{\prime}

\renewcommand\r{\rho}

\renewcommand\t{\tau}
\renewcommand\th{\theta}

\newcommand\wti{\widetilde}

\newcommand\cB{{\mathcal B}}

\newcommand\cE{{\mathcal E}}

\newcommand\cH{{\mathcal H}}

\newcommand\cV{{\mathcal V}}

\newcommand{\ct}[1]{\cite{#1}}
\newcommand{\bib}[1]{\bibitem{#1}}
\newcommand\PRL[3]{\textsl{Phys. Rev. Lett.} \textbf{#1}, #3 (#2)}

\newcommand\PRD[3]{\textsl{Phys. Rev.} \textbf{D#1}, #3 (#2)}

\newcommand\PLB[3]{\textsl{Phys. Lett.} \textbf{#1B}, #3 (#2)}

\newcommand\AoP[3]{\textsl{Ann. of Phys.} \textbf{#1}, #3 (#2)}

\newcommand\PR[3]{\textsl{Phys. Reports} \textbf{#1}, #3 (#2)}

\newcommand\IJMPA[3]{\textsl{Int. J. Mod. Phys.} \textbf{A#1}, #3 (#2)}

\newcommand\MPLA[3]{\textsl{Mod. Phys. Lett.} \textbf{A#1}, #3 (#2)}

\newcommand\phidot{\stackrel{.}{\phi}}

\newcommand\adot{\stackrel{.}{a}}

\newcommand\cBdot{\stackrel{.}{\mathcal B}}

\journalname{Eur. Phys. J. C}

\begin{document}

\title{Unified Dark Energy and Dust Dark Matter Dual to Quadratic Purely 
Kinetic K-Essence}
\titlerunning{Unified Dark Energy and Dust Dark Matter Dual to Quadratic Purely 
Kinetic K-Essence}

\author{Eduardo Guendelman\thanksref{e1,addr1}
        \and
        Emil Nissimov\thanksref{e2,addr2} 
	\and 
	Svetlana Pacheva\thanksref{e3,addr2} 
}

\thankstext{e1}{e-mail: guendel@bgu.ac.il}
\thankstext{e2}{e-mail: nissimov@inrne.bas.bg}
\thankstext{e3}{e-mail: svetlana@inrne.bas.bg}

\institute{Department of Physics, Ben-Gurion University of the Negev, Beer-Sheva, 
Israel \label{addr1}
          \and
Institute for Nuclear Research and Nuclear Energy,
Bulgarian Academy of Sciences, Sofia, Bulgaria \label{addr2}
}


\date{Received: date / Accepted: date}

\maketitle

\begin{abstract}

We consider a modified gravity plus single-scalar-field model, where the scalar
Lagrangian couples symmetrically both to the standard Riemannian volume-form
(spacetime integration measure density) given by the square-root of the determinant of
the Riemannian metric, as well as to another non-Riemannian volume-form 
in terms of an auxiliary maximal-rank antisymmetric tensor gauge field.
As shown in a previous paper, the pertinent
scalar field dynamics provides an exact unified description of both dark energy
via dynamical generation of a cosmological constant, and dark matter as a
``dust'' fluid with geodesic flow as a result of a hidden Noether symmetry.
Here we extend the discussion by considering a non-trivial modification of the 
purely gravitational action in the form of $f(R) = R - \a R^2$ generalized gravity.
Upon deriving the corresponding ``Einstein-frame'' effective action
of the latter modified gravity-scalar-field theory we find explicit duality 
(in the sense of weak versus strong coupling) between the original
model of unified dynamical dark energy and dust fluid dark matter, on one hand,
and a specific quadratic purely kinetic ``k-essence'' gravity-matter model
with special dependence of its coupling constants on only two independent parameters, 
on the other hand. The canonical Hamiltonian treatment and Wheeler-DeWitt quantization
of the dual purely kinetic ``k-essence'' gravity-matter model is also briefly discussed.



\end{abstract}

\section{Introduction}
\label{intro}



Unified description of dark energy and dark matter as a manifestation of a
single entity of matter has been an important challenge in cosmology in the 
last decade or so (for extensive reviews of dark energy see 
\ct{dark-energy-observ-1}-\ct{dark-energy-observ-3}, and for reviews of dark 
matter see \ct{dark-matter-rev-1}-\ct{dark-matter-rev-3}).

Originally, unified treatment of dark energy and dark matter was proposed in
the ``Chaplygin gas'' models \ct{chaplygin-1}-\ct{chaplygin-4}. Another trend
aimed at unifying dark energy and dark matter is based on the class of 
``k-essence'' models \ct{k-essence-1}-\ct{k-essence-4}, in particular, on
the so called ``purely kinetic k-essence'' models \ct{scherrer}
(for further developments, see \ct{pure-k-essence-1}-\ct{pure-k-essence-5}), 
which successfully avoid difficulties inherent in the generalized Chaplygin gas
models related to non-negligible sound speed.

Also, recently a lot of interest has been attracted by the so called 
``mimetic'' dark matter model proposed in \ct{mimetic-grav-1,mimetic-grav-2}. 
The latter employs a special covariant isolation of the conformal degree of 
freedom in Einstein gravity, whose dynamics mimics cold dark matter as a 
pressureless ``dust''. 
Further generalizations and extensions of ``mimetic'' gravity are studied in 
Refs.\ct{mimetic-grav-extend-1,mimetic-grav-extend-2}.

Models of explicitly coupled dark matter and dark energy described in terms
of two different scalar fields were proposed in Ref.\ct{copeland}.



In this paper we study a class of generalized models of gravity interacting
with a single scalar field employing the method of 
non-Riemannian volume-forms on the pertinent spacetime manifold, \textsl{i.e.},
generally covariant integration measure densities independent of the 
standard Riemannian one given in terms of the square root of the
determinant of the metric \ct{TMT-orig-0}-\ct{TMT-orig-4}.
(for further developments, see 
Ref.\ct{TMT-recent-2}). 
In this general class of models, also called
``two-measure gravity theories'', the non-Riemannian volume-forms are defined 
in terms of auxiliary maximal-rank antisymmetric tensor gauge fields 
(``measure gauge fields'').

The introduction of the two integration measures (one standard Riemannian
and another non-Riemannian one) opens the possibility to obtain both dark energy and 
dark matter from a single scalar field dynamics, as it was already observed in 
Ref.\ct{eduardo-singleton}.
Subsequently, in Ref.\ct{dusty} we have gone further and have discovered the 
fundamental reason that a class of models generalizing those studied in 
\ct{eduardo-singleton} 
describes a unification of dark matter and dark energy as an exact sum of 
two separate contributions in the pertinent energy-momentum tensor
(see also the review in Section 2 below). This is because of: 

(i) The appearance of an arbitrary integration constant from a dynamical
constraint on the scalar Lagrangian as a result of the equations of motion
for the auxiliary ``measure'' gauge field. This integration constant is
identified as a dynamically generated cosmological constant which provides 
the dark energy component.

(ii) The existence
of a hidden Noether symmetry of the non-Riemannian-measure-modified scalar
Lagrangian implying a conserved Noether current, which produces a dark matter
component as a ``dust'' fluid flowing along geodesics.

This behavior is totally independent of specific form of the scalar field Lagrangian, 
as long as the scalar field couples in a symmetric way to both of the measures. 
The latter also  ensures that the hidden ``dust'' Noether symmetry holds.
In addition, the fact that the dynamically generated cosmological constant 
arises as an arbitrary integration constant makes the observed vacuum energy density 
totally decoupled from the parameters of the initial scalar Lagrangian.

In the present paper our main object of study is a non-trivial extension of
the modified gravity plus single scalar field model considered in
\ct{dusty}, describing unification of dark matter and dark energy. 
Namely, we now modify the purely gravitational part of the action by introducing 
an extended $f(R)$-gravity action with $f(R) = R - \a R^2$ within the
first-order Palatini formalism
\foot{Let us recall that $R+R^2$ gravity within the second order formalism 
(which was also the first inflationary model) was originally
proposed in Ref.\ct{starobinsky}.}.

These new non-Riemannian\--measure\--modified gravity\--scalar\--field
theory has some very interesting features. On one hand, the hidden ``dust'' 
Noether symmetry of the scalar field Lagrangian remains intact and so does the
picture of unified description through the pertinent energy-momentum tensor of
both the dynamical dark energy (via the dynamical generation of the
cosmological constant) as well as the ``dust'' dark matter fluid,
\textsl{i.e.}, they appear as an exact sum of two separate contributions to
the corresponding energy density.

On the other hand, upon performing transition to the effective ``Einstein frame''
and upon appropriate scalar field redefinition we arrive at a {\em dual theory of a
specific quadratic purely kinetic ``k-essence''} form. 
In the dual purely kinetic ``k-essence''
formulation the original ``dust'' Noether symmetry is replaced by a simple
shift symmetry of the transformed scalar field. 

It is essential to stress on the following important properties of the dual
theory:

(a) All three constant coefficients in the quadratic purely kinetic 
``k-essence'' action are given in terms of only two independent parameters: $\a$
(the $R^2$ coupling constant) and an arbitrary integration constant $M$ produced by 
a dynamical constraint resulting from the equation of motion of the
``measure'' gauge field in the original theory.

(b) Let us emphasize that in the present context applying the notion of
``duality'' is justified in the usual sense of {\em weak coupling} in the
original theory versus {\em strong coupling} in the dual theory. Indeed, as we
will see below (Eqs.\rf{simplifying-notat}) all coupling constants in the dual
quadratic purely kinetic ``k-essence'' theory are functions of $1/\a$.

(c) In the limit $\a \to 0$ (strong coupling limit in the dual ``k-essence''
theory) the physical quantities -- energy density,
pressure {\em etc.} have well-defined smooth limiting values in spite of the
singularities in the coefficients of the ``k-essence'' action. The latter
limiting values coincide with the corresponding values in the original
gravity-matter theory with a standard Einstein-Hilbert gravity action
(where $\a=0$). Thus, the established duality reveals an important feature
of the {\em approximate} description of unified dark energy and dark matter 
via the quadratic purely kinetic ``k-essence'' model (Eq.\rf{k-essence-A} below)
-- it becomes an {\em exact} unification of both dark
species in the strong coupling limit $\a \to 0$.

(d) In the limit $\a \to 0$ the dual theory parameter 
$M$ precisely coincides with the dynamically generated effective cosmological
constant in the original theory \ct{dusty}. 
Thus, in this sense we can say that the dual purely kinetic
``k-essence'' theory is (partially) dynamically generated.

The above list of properties epitomizes the new features in our treatment of
the purely quadratic ``k-essence'' theory w.r.t. to the previous treatment in
\ct{scherrer}.


Now, an important remark is in order.
In a number of papers dealing with ``dust-like'' dark matter one uses the so called
``Lagrangian multiplier gravity'' formalism \ct{LMG-1,LMG-2}. We would like to point
out that this formalism is in fact a special particular case of the 
above mentioned more general and powerful approach based on non-Riemannian 
spacetime volume-forms (being used also in the present paper),
which appeared earlier and which has profound impact in any
(field theory) models with general coordinate reparametrization invariance, 
such as general relativity and its extensions \ct{TMT-orig-0}-\ct{TMT-orig-4}, 
\ct{TMT-recent-2}-\ct{dusty}, \ct{emergent}-\ct{buggy}; 
strings and (higher-dimensional) membranes 
\ct{mstring-2} and supergravity
\ct{susyssb-2}

Indeed, dynamical constraints like the one on the scalar field Lagrangian
in Eq.\rf{L-const} below, which routinely appear in all instances 
of applying the non-Riemannian volume-form method in gravity-matter theories,
resembles at first sight analogous constraints on scalar field Lagrangians in 
the ``Lagrangian multiplier gravity'' \ct{LMG-1,LMG-2}. However, employing
the concept of non-Riemannian volume-forms in the form of dual field
strengths of auxiliary maximal-rank tensor gauge fields (``measure gauge fields'') 
as in Eq.\rf{mod-measure} below, instead of bare Lagrange multiplier fields, 
has certain essential advantages:

\vspace{0.1in}

(i) Dynamical constraints in ``two-measure'' gravity-matter theories result
from the equations of motion of the auxiliary ``measure'' gauge fields and,
thus, they always involve an {\em arbitrary integration constant} like 
$M$ in Eq.\rf{L-const} below, as opposed to picking some {\em a priori} fixed constant
within the ``Lagrange multiplier gravity'' formalism. Depending on the
specific gravity-matter theories with non-Riemannian volume-forms under
consideration, the pertinent arbitrary integration constant acquires the
meaning of a {\em dynamically generated cosmological constant} like the
integration constant $M$ below 
(cf. Refs.\ct{TMT-orig-0,susyssb-2})

\vspace{0.1in}

(ii) Employing the canonical Hamiltonian formalism for Dirac-constrained
systems we find that the auxiliary ``measure'' gauge fields are in fact
almost pure gauge degrees of freedom except for the above mentioned
arbitrary integration constants which are identified with the conserved
Dirac-constrained canonical momenta conjugated to the ``magnetic'' components 
of the ``measure'' gauge fields 
(see Appendix A in Ref.\ct{buggy}). 

\vspace{0.1in}

(iii) Upon applying the non-Riemannian volume-form formalism to minimal $N=1$ 
supergravity the dynamically generated cosmological constant triggers spontaneous
supersymmetry breaking and mass generation for the gravitino (supersymmetric 
Brout-Englert-Higgs effect) 
\ct{susyssb-2}. Applying the same 
formalism to anti-de Sitter supergravity allows to produce
simultaneously a very large physical gravitino mass and a very small 
{\em positive} observable cosmological constant 
\ct{susyssb-2} in accordance 
with modern cosmological scenarios for slowly expanding universe of the present epoch
\ct{dark-energy-observ-1,dark-energy-observ-2,dark-energy-observ-3}. 

\vspace{0.1in}

(iv) Employing two independent non-Riemannian volume-forms in generalized
gravity-gauge+scalar-field models 
\ct{emergent}, 
thanks to the appearance of several arbitrary integration constants through the 
equations of motion w.r.t. the ``measure'' gauge fields, we obtain a
remarkable effective scalar potential with two infinitely large flat regions
(one for large negative and another one for large positive values of the scalar 
field $\vp$) with vastly different scales appropriate for a unified description 
of both the early and late universe' evolution.
An interesting feature is the existence of a stable initial phase of
{\em non-singular} universe creation preceding the inflationary phase
-- stable ``emergent universe'' without ``Big-Bang'' \ct{emergent}.
\vspace{0.1in}

The plan of the paper is as follows.
In Section 2, for self-consistency of the exposition, we briefly review
\ct{dusty}, namely, the basic aspects of the non-Riemannian volume-form approach 
leading to dynamical generation of a cosmological constant (dynamical dark energy) 
and revealing 
a hidden Noether symmetry allowing for a ``dust'' fluid representation of
dark matter, such that both dark species appear as a sum of two separate 
contributions to
the energy-momentum tensor. Sections 3 contains the main result -- establishing duality
(in the standard sense of weak versus strong coupling) between the original 
quadratic $f(R)$-gravity plus modified-measure scalar-field model, whose matter
part delivers an exact unified description of dynamical dark energy and dust 
fluid dark matter, on one hand, and a specific quadratic purely kinetic ``k-essence''
gravity-matter model, on the other hand. In Section 4 we analyze the above models
within the canonical Dirac-constrained Hamiltonian formalism and
derive the corresponding quantum Wheeler-DeWitt equations in the cosmological
Friedmann-Lemaitre-Robertson-Walker framework.

\section{Gravity-Matter Theory With a Non-Riemannian Volume-Form in the
Scalar Field Action -- Exact Unification of
Dark Energy and Dust Fluid Dark Matter}
\label{TMT}

\subsection{\textbf{Non-Riemannian Volume-Form Formalism}}
\label{TMT-formalism}

We start by considering the following non-conventional gravity-scalar-field
action -- a particular case of the general class of the so called ``two-measure''
gravity-matter theories \ct{TMT-orig-1}-\ct{TMT-orig-4}
(for simplicity we use units with the Newton constant $G_N = 1/16\pi$):
\be
S = S_{\rm grav} \lb g_{\m\n}, \G^\l_{\m\n}\rb + 
\int d^4 x \bigl(\sqrt{-g}+\P(B)\bigr) L(\vp,X) \; .
\lab{TMT}
\ee
The notations used here and below are as follows:
\begin{itemize}
\item
The first term in \rf{TMT} is the purely gravitational action in the first
order (Palatini) formalism, where the Riemannian metric $g_{\m\n}$ and the
affine connection $\G^\l_{\m\n}$ are {\em a priori} independent variables.
In the previous paper \ct{dusty} we have studied \rf{TMT} with the simplest
choice of $S_{\rm grav}$:
\be
S = \int d^4 x \sqrt{-g}\, R +
\int d^4 x \bigl(\sqrt{-g}+\P(B)\bigr) L(\vp,X) \; ,
\lab{TMT-0}
\ee
where $R$ denotes the scalar curvature -- in this case the first order
(Palatini) formalism is equivalent to the ordinary second-order (metric)
formalism.
\item
The second term in \rf{TMT} -- the scalar field action is constructed in
terms of two mutually independent volume-forms:

(a) $\sqrt{-g} \equiv \sqrt{-\det\Vert g_{\m\n}\Vert}$ is the standard
Riemannian integration measure density (spacetime volume-form);

(b) $\P(B)$ denotes an alternative non-Riemannian generally covariant
integration measure density defining an alternative non-Riemannian volume-form:
\be
\P(B) = \frac{1}{3!}\vareps^{\m\n\k\l} \pa_\m B_{\n\k\l} \; ,
\lab{mod-measure}
\ee
where $B_{\m\n\l}$ is an auxiliary maximal rank antisymmetric tensor gauge
field independent of the Riemannian metric. 

$B_{\m\n\l}$ \rf{mod-measure} is called ``measure gauge field''.
\item
$L(\vp,X)$ is general-coordinate invariant Lagrangian of a single scalar field 
$\vp (x)$, which can be of an arbitrary generic ``k-essence'' form 
\ct{k-essence-1,k-essence-2,k-essence-3,k-essence-4}:
\br
L(\vp,X) = \sum_{n=1}^N A_n (\vp) X^n - V(\vp) \; ,
\lab{k-essence-L} \\
X \equiv - \h g^{\m\n}\pa_\m \vp \pa_\n \vp \; ,
\nonu
\er
\textsl{i.e.}, a nonlinear (in general) function of the scalar kinetic term $X$.
\end{itemize}

In this Section we will concentrate on the scalar field action -- the second
term in \rf{TMT}. First, due to general-coordinate invariance we have
covariant conservation of the pertinent energy-momentum tensor:
\br
T_{\m\n} = g_{\m\n} L(\vp,X) + 
\Bigl( 1+\frac{\P(B)}{\sqrt{-g}}\Bigr) \partder{L}{X} \pa_\m \vp\, \pa_\n \vp \; ,
\lab{EM-tensor} \\
\nabla^\n T_{\m\n} = 0 \quad\quad \; .
\nonu
\er
It follows from the equation of motion w.r.t. $\vp$:
\be
\partder{L}{\vp} + \bigl(\P(B)+\sqrt{-g}\bigr)^{-1}
\pa_\m \Bigl\lb\bigl(\P(B)+\sqrt{-g}\bigr) 
g^{\m\n}\pa_\n \vp \partder{L}{X}\Bigr\rb = 0 \; ;
\lab{vp-eqs}
\ee
Further, variation of the action \rf{TMT} w.r.t. ``measure'' gauge field $B_{\m\n\l}$
reads:
\be
\pa_\m L (\vp,X) = 0 \; ,
\lab{L-eq}
\ee
\textsl{i.e.}, the $B_{\m\n\l}$-equations of motion yield the following
dynamical constraint on the scalar field Lagrangian:
\be
L (\vp,X) = - 2M = {\rm const} \; ,
\lab{L-const}
\ee
where $M$ is arbitrary integration constant. The factor $2$ in front of $M$ is for later
convenience, moreover, we will take $M>0$ in view of its interpretation as
a dynamically generated cosmological constant (see Eqs.\rf{EM-tensor-0} and
\rf{T-J-hydro}, \rf{p-const} below).

It is important to stress that the scalar field dynamics is determined entirely by 
the first-order differential equation - the dynamical constraint
Eq.\rf{L-const} which, in the simplest case of \rf{k-essence-L}
$(L(\vp,X)=X-V(\vp)$) to be considered henceforth for simplicity, implies:
\be
X - V(\vp) = - 2M \;\; \longrightarrow \;\; X = V(\vp) - 2M \; .
\lab{X-constr}
\ee
The standard second order differential equation \rf{vp-eqs} is in fact a consequence
of \rf{L-const} together with the energy-momentum conservation 
$\nabla^\m T_{\m\n} = 0$ with:
\be
T_{\m\n} = - 2M g_{\m\n} + 
\Bigl( 1+\frac{\P(B)}{\sqrt{-g}}\Bigr)\pa_\m \vp \pa_\n \vp \; .
\lab{EM-tensor-0}
\ee

The physical meaning of the ``measure'' gauge field $B_{\m\n\l}$ \rf{mod-measure}
as well as the meaning of the integration constant  $M$ are most straightforwardly 
seen within the canonical Hamiltonian treatment of (the scalar field part of) \rf{TMT} 
-- this is systematically derived in Sec.2 of Ref.\ct{dusty}. Namely, using
short-hand notations for the components of $B_{\m\n\l}$ \rf{mod-measure}:
\br
\P(B) = \pa_\m \cB^\m = \cBdot + \pa_i \cB^i \;\; ,\;\;
\cB^\m \equiv \frac{1}{3!}\vareps^{\m\n\k\l} B_{\n\k\l} \; ,
\lab{B-short-0} \\
\cB \equiv\cB^0 = \frac{1}{3!} \vareps^{mkl} B_{mkl} \;\; ,\;\;
\cB^i \equiv -\h \vareps^{ikl} B_{0kl} \; ,
\lab{B-short}
\er
we obtain for the canonically conjugated momenta $\pi_{\cB},\pi_{\cB^i}$ the
set of Dirac first class constraints:
\be
\pi_{\cB^i} = 0 \quad, \quad
\pa_i \pi_{\cB} = 0 \quad \longrightarrow \quad \pi_{\cB} = {\rm const}\equiv -2 M \; ,
\lab{M-constr}
\ee
where we have $\pi_{\cB} = L(\vp,X)$ straightforwardly obtainable from the
explicit form of the action \rf{TMT} taking into account the representation of
$\P(B)$ in \rf{B-short-0}.
The above Dirac constraints imply that all components of the ``measure'' gauge 
field $B_{\m\n\l}$ \rf{B-short} are pure-gauge (non-propagating) degrees of freedom. 
The last relation in \rf{M-constr} is the canonical Hamiltonian analog of 
Eq.\rf{L-const} within the Lagrangian formalism -- in other words, the
integration constant $M$ in the $B_{\m\n\l}$-equations of motion is (modulo
a trivial numerical factor) a conserved Dirac-constrained canonical momentum 
$\pi_{\cB}$ conjugated to the ``magnetic'' component $\cB$ of the
``measure''-gauge field $B_{\m\n\l}$ \rf{B-short}. 

For more details about the canonical Hamiltonian treatment of general 
gravity-matter theories with (several independent) non-Riemannian volume-forms 
we refer to 
\ct{buggy}.

\subsection{\textbf{``Dust'' Fluid Conservation Laws}}
\label{conserv-law}

In Ref.\ct{dusty} we have shown that the scalar field action in \rf{TMT}
possesses a hidden Noether symmetry, namely
\rf{TMT} is invariant (up to a total derivative) under the
following nonlinear symmetry transformations:
\br
\d_\eps \vp = \eps \sqrt{X} \quad ,\quad \d_\eps g_{\m\n} = 0 \; ,
\nonu \\
\d_\eps B_{\m\n\l} = - \eps \frac{1}{2\sqrt{X}} \vareps_{\m\n\l\k}
g^{\k\r}\pa_\r \vp \, \bigl(\P(B) + \sqrt{-g}\bigr)  \; .
\lab{hidden-sym}
\er
Then, the standard Noether procedure yields the conserved current:
\be
\nabla_\m J^\m = 0 \quad ,\quad
J^\m \equiv \Bigl(1+\frac{\P(B)}{\sqrt{-g}}\Bigr)\sqrt{2X} 
g^{\m\n}\pa_\n \vp \partder{L}{X} \; .
\lab{J-conserv}
\ee

Let us stress at this point, that the existence of the hidden symmetry
\rf{hidden-sym} of the action \rf{TMT} {\em does not} depend on the specific
form of the scalar field Lagrangian \rf{k-essence-L}. The only requirement is 
that the kinetic term $X$ must be positive.

$T_{\m\n}$ \rf{EM-tensor} and $J^\m$ \rf{J-conserv} can be cast into
the a relativistic hydrodynamical form (taking into account \rf{L-const}):
\be
T_{\m\n} = - 2M g_{\m\n} + \rho_0 u_\m u_\n \quad ,\quad J^\m = \rho_0 u^\m \; ,
\lab{T-J-hydro}
\ee
where:
\br
\rho_0 \equiv \Bigl(1+\frac{\P(B)}{\sqrt{-g}}\Bigr)\, 2X \partder{L}{X} \; ,
\lab{rho-0-def} \\
u_\m \equiv \frac{\pa_\m \vp}{\sqrt{2X}} \quad 
({\rm note} \; u^\m u_\m = -1\;) \; .
\lab{u-def}
\er
For the pressure $p$ and energy density $\rho$ we have accordingly:
\br
p = - 2M = {\rm const} \; ,
\lab{p-const} \\
\rho = \rho_0 - p = \Bigl(1+\frac{\P(B)}{\sqrt{-g}}\Bigr)\, 2X \partder{L}{X}
+ 2M \; ,
\lab{rho-def}
\er
where the integration constant $M$ 
appears as {\em dynamically generated cosmological constant}.

Let us note that constancy \rf{p-const} of the pressure $p=-2M$ in \rf{T-J-hydro}
together with the covariant conservation of $T_{\m\n}$:
\br
\nabla^\n T_{\m\n} = \nabla^\n \bigl(\rho_0 u_\m u_\n\bigr) 
\nonu \\
= u_\m \nabla^\n \bigl(\rho_0 u_\n\bigr) + 
\rho_0 \bigl( u_\n \nabla^\n u_\m \bigr) = 0 \; ,
\lab{T-conserv}
\er
upon projecting \rf{T-conserv} along the ``velocity'' vector $u_\m$
and orthogonally w.r.t. the latter by $\Pi^{\m\l}=g^{\m\l}+u^\m u^\l$),
immediately implies {\em both the covariant conservation of} $J^\m$
\rf{J-conserv}:
\be
\nabla_\m \bigl(\rho_0 u^\m\bigr) = 0 \; ,
\lab{dust-conserv}
\ee
{\em as well as the geodesic flow equation}:
\be
u_\n \nabla^\n u_\m = 0
\lab{geodesic-eq}
\ee

As discussed in \ct{dusty} the energy-momentum tensor \rf{T-J-hydro} consists of two 
parts with the following interpretation according to the standard $\L$-CDM model
\ct{Lambda-CDM-1,Lambda-CDM-2,Lambda-CDM-3} (using notations
$p = p_{\rm DM} + p_{\rm DE}$ and $\rho = \rho_{\rm DM} + \rho_{\rm DE}$ in
\rf{p-const}-\rf{rho-def}):
\begin{itemize}
\item
Dark energy part given by the first  cosmological constant term in $T_{\m\n}$
\rf{T-J-hydro}, which arises due to the dynamical constraint on the scalar field
Lagrangian \rf{L-const} 
with $p_{\rm DE} = -2M\, ,\, \rho_{\rm DE} = 2M$;
\item
Dark matter part given by the second term in \rf{T-J-hydro}
with $p_{\rm DM} = 0\, ,\, \rho_{\rm DM} = \rho_0$
($\rho_0$ as in \rf{rho-0-def}), which in fact describes a {\em dust} fluid.
According to the general definitions, see \textsl{i.e.} \ct{rezzola-zanotti},
$\rho_{\rm DM} = \rho_0$ \rf{rho-0-def} and $\rho_{\rm DE} = 2M$ are the rest-mass and 
internal fluid energy densities, so that the Noether conservation law \rf{J-conserv}
describes dust dark matter ``particle number'' conservation.
\end{itemize}

\section{Quadratic Gravity Interacting with a Dark Energy - Dark Matter
Unifying Scalar field  -- Dual to Purely Kinetic ``K-Essence''}
\label{duality}

\subsection{\textbf{Derivation of the Dual Kinetic Pure ``K-Essence'' Theory}}
\label{duality-A}

Let us now consider a modification of the gravitational part of the 
gravity-scalar-field action \rf{TMT-0} as follows:
\br
S = \int d^4 x \sqrt{-g} \bigl( R(g,\G) - \a R^2(g,\G)\bigr)
\nonu \\
+ \int d^4 x \bigl(\sqrt{-g}+\P(B)\bigr) L(\vp,X) \; ,
\lab{TMT-A}
\er
where we have introduced $f(R)=R-\a R^2$ extended gravity action in the 
first-order Palatini formalism:
\be
R(g,\G) = g^{\m\n} R_{\m\n}(\G) \; ,
\lab{palatini}
\ee
\textsl{i.e.}, with {\em a priori} independent metric $g_{\m\n}$ and affine 
connection $\G^\m_{\n\l}$.

Clearly, since the scalar field action -- the second term in \rf{TMT-A} --
remains the same as in the original action \rf{TMT-0}, all results in
subsection 2.2 remain valid
\foot{Adding bare cosmological constant term $-2\L_0 \sqrt{-g}$ to the
gravity action in \rf{TMT-A} is irrelevant, since this is equivalent to a
constant shift of the scalar Lagrangian $L(\vp,X) \to L(\vp,X) - 2 \L_0$
(recall that the non-Riemannian measure density $\P(B)$ is total derivative
\rf{mod-measure}), which in turn amounts on-shell to a trivial redefinition of the
arbitrary integration constant $M$ ($M \to M+\L_0$) in Eq.\rf{L-const}.}
. 
In other words, the modified gravity-scalar-field action \rf{TMT-A} possesses 
``hidden'' Noether symmetry \rf{hidden-sym} producing ``dust'' fluid energy
density conserved current \rf{dust-conserv} and interpretation of $\vp$ as 
describing simultaneously dark energy and dust dark matter with geodesic dust 
fluid flow \rf{geodesic-eq}.

The gravitational equations of motion resulting from \rf{TMT-A} are,
however, not of the standard Einstein form. The equations of motion w.r.t.
$g^{\m\n}$ read:
\br
R_{\m\n}(\G) = \frac{1}{2 f^{\pr}_R} \bigl\lb T_{\m\n} + f(R) g_{\m\n} \bigr\rb \; ,
\phantom{aaaaa}
\lab{non-einstein-eqs} \\
f(R) = R(g,\G ) - \a R^2(g,\G ) \; ,\; f^{\pr}_R = 1 - 2\a R(g,\G ) \; ,
\lab{f-R-notat}
\er
with $T_{\m\n}$ the same as in \rf{EM-tensor-0}. Trace of Eqs.\rf{non-einstein-eqs}
yields:
\br
R(g,\G)= -\h T \quad ,\quad T = g^{\m\n} T_{\m\n} \; ,
\lab{non-einstein-eqs-trace} \\
T = - 8M - \Bigl( 1+\frac{\P(B)}{\sqrt{-g}}\Bigr)\,2\bigl(V(\vp) - 2M\bigr) \; .
\lab{EM-tensor-0-trace}
\er

Variation of \rf{TMT-A} w.r.t. $\G^\m_{\n\l}$ yields:
\be
\int d^4\,x\,\sqrt{-g} g^{\m\n} f^{\pr}_R
\(\nabla_\k \d\G^\k_{\m\n} - \nabla_\m \d\G^\k_{\k\n}\) = 0 
\lab{var-G}
\ee
which shows, following the analogous derivation in the Ref.\ct{TMT-orig-1}, that 
$\G^\m_{\n\l}$ becomes a Levi-Civita connection:
\be
\G^\m_{\n\l} = \G^\m_{\n\l}({\overline g}) = 
\h {\overline g}^{\m\k}\(\pa_\n {\overline g}_{\l\k} + \pa_\l {\overline g}_{\n\k} 
- \pa_\k {\overline g}_{\n\l}\) \; ,
\lab{G-eq}
\ee
w.r.t. to the Weyl-rescaled metric ${\overline g}_{\m\n}$:
\be
{\overline g}_{\m\n} = f^{\pr}_R\, g_{\m\n} \; .
\lab{bar-g}
\ee

Before going over to the physical ``Einstein-frame'' it is useful to perform
the following $\vp$-field redefinition:
\br
\vp \to {\wti \vp} = \int \frac{d\vp}{\sqrt{\bigl( V(\vp)-2M\bigr)}} \; ,
\lab{new-vp} \\
X \to {\wti X} = -\h {\overline g}^{\m\n} \pa_\m {\wti \vp} \pa_\n {\wti \vp}
= \frac{1}{f^{\pr}_R} \; ,
\lab{wti-X-eq}
\er
where the last relation follows from the Lagrangian dynamical constraint \rf{X-constr} 
together with \rf{bar-g}.

Now, using Eqs.\rf{f-R-notat}-\rf{EM-tensor-0-trace}, which together with \rf{wti-X-eq}
imply:
\be
\frac{1}{\wti X} = 1 - \a \Bigl\lb 8M +
\Bigl( 1+\frac{\P(B)}{\sqrt{-g}}\Bigr)\,2\bigl(V(\vp) - 2M\bigr)\Bigr\rb,
\lab{wti-X-eq-1}
\ee
as well as Eqs.\rf{G-eq}-\rf{bar-g} and \rf{wti-X-eq} we can rewrite all equations
of motion resulting from \rf{TMT-A}, in particular the quadratic $f(R)$-gravity 
Eqs.\rf{non-einstein-eqs},
in terms of the new  metric ${\overline g}_{\m\n}$ \rf{bar-g} and the new
scalar field ${\wti \vp}$ \rf{new-vp} in the standard form of Einstein
gravity equations:
\be
{\overline R}_{\m\n} - \h {\overline g}_{\m\n} {\overline R} = 
\h {\overline T}_{\m\n}
\lab{standard-einstein-eqs}
\ee
with the following notations:
\begin{itemize}
\item
Here ${\overline R}_{\m\n}$ and ${\overline R}$ are the standard Ricci
tensor and scalar curvature of the Einstein-frame metric \rf{bar-g}.
\item
The Einstein-frame energy-momentum tensor:
\be
{\overline T}_{\m\n} = {\overline g}_{\m\n} {\overline L}_{\rm eff}
- 2 \partder{{\overline L}_{\rm eff}}{{\overline g}^{\m\n}}
\lab{EM-tensor-eff}
\ee
is given in terms of the following effective ${\wti \vp}$-scalar field Lagrangian 
of a specific quadratic purely kinetic ``k-essence'' form:
\br
{\overline L}_{\rm eff} ({\wti X}) = A_2 {\wti X}^2 - A_1 {\wti X} + A_0
\lab{L-eff} \\
A_2 \equiv \frac{1}{4\a} - 2M \;\; ,\;\; A_1 \equiv \frac{1}{2\a} 
,\;\; A_0 \equiv \frac{1}{4\a} \; . 
\lab{simplifying-notat}
\er
\end{itemize}
Let us stress that the three constant coefficients in \rf{L-eff} depend only
on two independent parameters $(\a, M)$, the second one being a dynamically generated
integration constant in the original theory \rf{TMT-A}.

Thus, we have established a {\em duality} between the modified-measure
gravity-scalar-field theory \rf{TMT-A} within the original $g_{\m\n}$-frame
and the special quadratic purely kinetic ``k-essence'' theory within the
conformally-rescaled ${\overline g_{\m\n}}$-frame (Einstein-frame):
\be
S_{\rm k-ess} = \int d^4 \sqrt{-{\overline g}} \Bigl\lb {\overline R} +
\Bigl(\frac{1}{4\a} - 2M\Bigr) {\wti X}^2 - \frac{1}{2\a} {\wti X} + 
\frac{1}{4\a}\Bigr\rb \; .
\lab{k-essence-A}
\ee
with a matter Lagrangian \rf{L-eff}-\rf{simplifying-notat}.

The Einstein-frame effective energy-momentum-tensor \rf{EM-tensor-eff} 
in the perfect fluid representation reads (taking into account 
\rf{L-eff}-\rf{simplifying-notat}):
\br
{\bar T}_{\m\n} = {\overline g}_{\m\n} {\wti p} + 
{\wti u}_\m {\wti u}_\n \bigl({\wti \rho} + {\wti p}\bigr) \; ,
\lab{EM-tensor-wti} \\
{\wti p} = \Bigl(\frac{1}{4\a} - 2M\Bigr) {\wti X}^2  
- \frac{1}{2\a} {\wti X} + \frac{1}{4\a} \; ,
\lab{wti-p} \\
{\wti \rho} = 3 \Bigl(\frac{1}{4\a} - 2M\Bigr){\wti X}^2 
- \frac{1}{2\a} {\wti X} - \frac{1}{4\a} \; , 
\lab{wti-rho} \\
{\wti u}_\m \equiv \frac{\pa_\m {\wti \vp}}{\sqrt{2{\wti X}}} \;\; , \;\;
{\overline g}^{\m\n} {\wti u}_\m {\wti u}_\n = - 1 \; .
\nonu
\er

Because of the obvious Noether symmetry of \rf{k-essence-A} under
constant shift of ${\wti \vp}$:
\be
{\wti \vp} \to {\wti \vp} + {\rm const}
\lab{shift-sym}
\ee
the corresponding
Noether conservation law is identical to the ${\wti \vp}$-equations of motion:
\be
{\overline \nabla}_\m \Bigl({\overline g}^{\m\n} \pa_\n {\wti \vp}
\partder{{\wti L}_{\rm eff}}{{\wti X}}\Bigr) = 0 \; ,
\lab{vp-eqs-wti}
\ee
where ${\overline \nabla}_\m$ indicates covariant derivative w.r.t.
Levi-Civita connection in the ${\overline g}_{\m\n}$-(Einstein) frame.


Introducing now the standard thermodynamical notions of enthalpy per unit
particle ${\wti h}$ and particle number density ${\wti n}$ within the above
Einstein-frame system, where ${\wti \rho} + {\wti p} = {\wti h}{\wti n}$,
and taking into account the expressions \rf{wti-p}-\rf{wti-rho}, the Noether current
conservation law \rf{vp-eqs-wti} and the covariant conservation 
${\overline \nabla}^\n {\wti T}_{\m\n} = 0$ of \rf{EM-tensor-wti} can be
written, respectively, as:
\be
{\overline \nabla}_\m \bigl({\wti n}{\wti u}^\m \bigr) = 0 \;\; ,\;\;
{\wti u}^\n {\overline \nabla}_\n {\wti u}^\m +
{\wti \Pi}_\m^\n \pa_\n \ln {\wti h} = 0 \; ,
\lab{no-geodesic}
\ee
where:
\br
{\wti h} = \sqrt{2{\wti X}} \quad ,\quad
{\wti \Pi}^{\m\n} = {\overline g}^{\m\n} + {\wti u}^\m {\wti u}^\n \; ,
\nonu \\
{\wti n} = \sqrt{2{\wti X}} \partder{{\wti L}_{\rm eff}}{{\wti X}}
= \sqrt{2{\wti X}} \Bigl\lb \frac{1}{2\a}\bigl({\wti X}-1\bigr) 
- 4M{\wti X}\Bigr\rb\; .
\lab{wti-notation}
\er

Comparing relations \rf{no-geodesic} in the Einstein frame with the corresponding 
relations in the original $g_{\m\n}$-frame \rf{dust-conserv}-\rf{geodesic-eq} 
we conclude that:
\begin{itemize}
\item
Conservation of the dust dark matter energy density current
\rf{dust-conserv} in the $g_{\m\n}$-frame is dual to the conservation of the
particle number density current within the Einstein frame
--  first Eq.\rf{no-geodesic}, which in fact is the standard 
${\wti\vp}$-equation of motion resulting from the action \rf{k-essence-A}.
That is, the ``hidden'' nonlinear Noether symmetry of \rf{TMT-A}
is dual to the shift-symmetry \rf{shift-sym} of \rf{k-essence-A}. 
\item
In the original $g_{\m\n}$-frame the dust dark matter flows along geodesics
\rf{geodesic-eq}, whereas in the dual Einstein frame the dual purely kinetic
``k-essence'' fluid does not any more flow along geodesics (second Eq.\rf{no-geodesic}).
\end{itemize}

The purely kinetic ``k-essence'' theory \rf{k-essence-A} apart from the
trivial vacuums ${\wti \vp} = {\rm const}$ possesses in addition a non-trivial
``kinetic vacuum'' solution ${\wti X}_{\rm vac}$ of the equations of motion
\rf{vp-eqs-wti}:
\be
\partder{{\wti L}_{\rm eff}}{{\wti X}}\bgv_{{\wti X}_{\rm vac}} = 0
\;\; \longrightarrow \;\;
{\wti X}_{\rm vac} = 
\frac{1}{1- 8\a M} \; ,
\lab{kin-vac}
\ee
implying the dark energy property (cf. \rf{EM-tensor-wti}):
\be
\bigl({\wti \rho} + {\wti p}\bigr)\bv_{{\wti X}_{\rm vac}} = 0 \;\; ,\;\;
{\wti \rho}\bv_{{\wti X}_{\rm vac}} = 
\frac{2M}{1- 8\a M} \equiv 2\L_{\rm eff}
\lab{CC-kin-vac}
\ee
with effective cosmological constant:
\be
\L_{\rm eff} = 
\frac{M}{1- 8\a M} \; .
\lab{CC-eff}
\ee
The explicit form of \rf{kin-vac} reads:
\be
{\overline g}^{\m\n} \pa_\m {\wti \vp}_{\rm vac} \pa_\n {\wti \vp}_{\rm vac} 
+ \frac{2}{1- 8\a M} = 0 \; .
\lab{ham-jacob}
\ee
It has the form of the standard Hamilton-Jacobi equation for the action of a massive
relativistic point-particle moving in a ${\overline g}_{\m\n}$-background
with mass squared:
\be
m_0^2 
\equiv \frac{2}{1- 8\a M} \; .
\lab{m-0}
\ee
In other words ${\wti \vp}_{\rm vac}(x) = m_0 T$, where $T$ is the
proper-time for the particle to reach the spacetime point $x$ from some
reference point $x_{(0)}$.

\subsection{\textbf{FLRW Reduction of the Dual Kinetic Pure ``K-Essence'' Theory}}
\label{FLRW}

Let us now consider a reduction of the dual quadratic purely kinetic ``k-essence'' 
gravity-scalar-field model \rf{k-essence-A} for the
Friedman-Lemaitre-Robertson-Walker (FLRW) class of metrics:
\be
ds^2 = - N^2(t) dt^2 + a^2(t) \Bigl\lb \frac{dr^2}{1-K r^2}
+ r^2 (d\th^2 + \sin^2\!\th d\phi^2)\Bigr\rb \; .
\lab{FLRW}
\ee
Then the action \rf{k-essence-A} acquires the form (using again short-hand notations 
\rf{simplifying-notat}; in what follows we will take the spatial FLRW curvature $K=0$ 
for simplicity):
\be
S = \int dt \Bigl\lb - \frac{a\,\adot^2}{N} + 
\frac{N}{4}a^3 \Bigl( \frac{1}{\a} - \frac{1}{\a} \frac{\phidot^2}{N^2} +
\bigl(\frac{1}{4\a}-2M\bigr) \frac{\phidot^4}{N^4}\Bigr)\Bigr\rb \; ,
\lab{FLRW-action-A}
\ee
with $\phidot \equiv \frac{d{\wti \vp}}{dt}$.
The $\phi \equiv {\wti \vp}$-equation of motion from \rf{FLRW-action-A} yields 
(using henceforth the gauge $N=1$):
\be
\frac{d p_\phi}{dt} = 0 \; \longrightarrow \; 
p_\phi = a^3 \Bigl\lb - \frac{1}{2\a} \phidot + 
\bigl(\frac{1}{4\a} - 2M\bigr) \phidot^3 \Bigr\rb \; ,
\lab{phi-eq}
\ee
where $p_\phi$ is the constant conserved canonically conjugated momentum 
of $\phi\equiv {\wti \vp}$.

The equation of motion w.r.t. $a$ resulting from \rf{FLRW-action-A} -- the
Friedmann equation -- reads:
\be
\adot^2 = \frac{1}{6} a^2 \rho (p_\phi/a^3) \; ,
\lab{friedman-eq}
\ee
with $\rho (p_\phi/a^3)$ denoting the energy density as function of
$\frac{p_\phi}{a^3}$:
\br 
\rho (p_\phi/a^3) = 
- A_0 -\h A_1 \phidot^2\!\!(p_\phi/a^3) 
+ \frac{3}{4} A_2 \phidot^4\!\!(p_\phi/a^3) 
\nonu \\
= \frac{1}{8\a} \phidot^2\!\!(p_\phi/a^3) +
\frac{3}{4}\frac{p_\phi}{a^3}\!\!\phidot (p_\phi/a^3) - \frac{1}{4\a} \; ,
\phantom{aaa}
\lab{rho-A}
\er
where in the second line of \rf{rho-A} relation \rf{phi-eq} was used.
Here $\phidot\!\!(p_\phi/a^3)\! \equiv\! y$ is one of the roots of the cubic equation 
\rf{phi-eq}:
\be
y^3 - \frac{2}{1-8\a M} y - \frac{4\a}{1-8\a M} \frac{p_\phi}{a^3} = 0 \; ,
\lab{cubic-eq}
\ee
which explicitly read:
\br
y_1 = {\wti A} + {\wti B} \;\; , \;\;
y_{2,3}= - \h \bigl({\wti A} + {\wti B}\bigr) \pm 
i \frac{\sqrt{3}}{2} \bigl({\wti A} - {\wti B}\bigr) \; ,
\lab{cubic-root}
\er
where:
\br
{\wti A}, {\wti B}\equiv \frac{2\a}{1-8\a M} \frac{p_\phi}{a^3} 
\phantom{aaaaa}
\nonu \\
 \pm \sqrt{\Bigl(\frac{2\a}{1-8\a M} \frac{p_\phi}{ a^3}\Bigr)^2 - 
\Bigl(\frac{2}{3(1-8\a M)}\Bigr)^3}
\lab{cubic-notat}
\er

Similarly, for the pressure we have:
\br
p (p_\phi/a^3) = A_0 -\h A_1 \phidot^2\!\!(p_\phi/a^3) 
+ \frac{1}{4} A_2 \phidot^4\!\!(p_\phi/a^3) 
\nonu \\
= - \frac{1}{8\a} \phidot^2\!\!(p_\phi/a^3) +
\frac{p_\phi}{4a^3} \phidot\!\!(p_\phi/a^3) + \frac{1}{4\a} \; ,
\lab{p-A}
\er
where again in the second line the cubic equation \rf{cubic-eq} has been used.

The Friedmann equation \rf{friedman-eq} can be solved approximately for
small and large values of $a$ using expressions \rf{cubic-root}-\rf{cubic-notat}:
\br
\phidot\!\!(p_\phi/a^3) \simeq \Bigl(\frac{2\a p_\phi}{1-8\a M}\Bigr)^{1/3} a^{-1} 
\quad {\rm for} \; a \to 0 \; ,
\lab{y-small-a} \\
\phidot\!\!(p_\phi/a^3) \simeq \sqrt{\frac{2}{1-8\a M}}
+ \a \,\frac{p_\phi}{a^3}
\quad {\rm for} \; a \to \infty \; .
\lab{y-large-a}
\er
Accordingly, we have for the energy density \rf{rho-A}:
\be
\rho (p_\phi/a^3) \simeq \frac{3}{4} \Bigl(\frac{4\a}{1-8\a M}\Bigr)^{1/3}
p_\phi^{4/3} a^{-4} \quad
{\rm for} \; a \to 0 \; ,
\lab{rho-small-a} 
\ee
\textsl{i.e.} radiation domination for small $a$, and
(using notations \rf{CC-eff}, \rf{m-0}):
\br
\rho (p_\phi/a^3) \simeq \frac{2M}{1-8\a M} 
+ \sqrt{\frac{2}{1-8\a M}}\;\frac{p_\phi}{a^3}
\lab{rho-large-a} \\
=  2 \L_{\rm eff} +  m_0 \,\frac{p_\phi}{a^3} \quad
{\rm for} \; a \to \infty \; ,
\lab{rho-large-b}
\er
\textsl{i.e.}, dark energy domination plus a subleading ``dust'' dark matter
contribution for large $a$. 


An important property of the above derivation of the duality between the
original $g_{\m\n}$-frame quadratic $f(R)$-gravity plus non-Riemannian-modified-measure 
scalar field action \rf{TMT-A}, on one hand, and the Einstein-frame special
quadratic purely kinetic ``k-essence'' theory \rf{k-essence-A}, on the other hand, is
that there exists a smooth limit $\a \to 0$ of  
the energy density \rf{rho-A} and pressure \rf{p-A}
in the latter theory in spite of the singularity at the strong coupling limit
$\a \to 0$ in all kinetic ``k-essence'' coefficients \rf{L-eff}-\rf{simplifying-notat}.
The corresponding limiting values at $\a=0$ of the energy density and
pressure of the dual purely kinetic ``k-essence'' theory \rf{k-essence-A}
are those of the original theory \rf{TMT-A} with the standard 
Einstein gravity action ($\a=0$), \textsl{i.e.}, the theory \rf{TMT-0} \ct{dusty}.
In particular, in the limit $\a \to 0$ we get precisely the expression for the 
energy density being an {\em exact sum} of dark energy and dust dark matter
contributions produced by \rf{TMT-0} \ct{dusty}.

Indeed, from \rf{phi-eq} and \rf{cubic-root}-\rf{cubic-notat}
we obtain for $\a \to 0$:
\br
\phidot (p_\phi/a^3) \simeq \sqrt{2} 
+ \a \Bigl(4\sqrt{2}M + \frac{p_\phi}{a^3}\Bigr)
\nonu \\
+ \a^2 \Bigl\lb 24\sqrt{2}M - 
\frac{3\sqrt{2}}{4} \bigl(\frac{p_\phi}{a^3}\bigr)^2\Bigr\rb 
+ {\rm O}(\a^3) \; ,
\lab{phidot-small-a}
\er
wherefrom \rf{rho-A} and \rf{p-A} yield for small $\a$:
\br
\rho (p_\phi/a^3) \simeq 2M + \sqrt{2}\frac{p_\phi}{a^3}  
\nonu \\
+ \a \Bigl\lb 16M^2 + 4\sqrt{2}M \frac{p_\phi}{a^3} + 
\h \bigl(\frac{p_\phi}{a^3}\bigr)^2 \Bigr\rb + {\rm O}(\a^2) \; ,
\lab{rho-small-alpha} \\
p (p_\phi/a^3) \simeq - 2M - \a \Bigl\lb 16M^2 - 
\h \bigl(\frac{p_\phi}{a^3}\bigr)^2 \Bigr\rb  + {\rm O}(\a^2) \; .
\lab{p-small-alpha}
\er

The expression \rf{rho-small-alpha} resembles the large-$a$
asymptotics \rf{rho-large-a} for the energy density, however, unlike the
latter Eq.\rf{rho-small-alpha} is valid for generic values of the FLRW factor
$a$ (and small values of $\a$ -- the $R^2$ coupling constant in the original
theory \rf{TMT-A}).

The cosmological implications of general purely kinetic ``k-essence'' models have been 
previously studied extensively in Refs.\ct{scherrer}-\ct{pure-k-essence-5}.
In particular, an important strong inequality (Eqs.(26)-(27) in \ct{scherrer}) 
involving the parameters of a generic quadratic purely kinetic ``k-essence'' theory 
was derived from the requirement that the onset of dark matter behavior must
occur before the epoch of equal matter and radiation, and also
that in the present stage of evolution the dark energy component
must exceed twice the dark matter component in the pertinent energy density,
The counterpart of the above Scherrer's inequality in the present settings
becomes the requirement:
\be
\a M \ll 1 \; ,
\lab{scherrer}
\ee
which completely conforms to the result
\rf{rho-small-alpha}.

From Eqs.\rf{rho-A}-\rf{p-A} we obtain for the squared sound speed:
\be
c_s^2 = \partder{p}{\rho} = \frac{1}{3}\Bigl\lb 1 - 
\frac{1}{\sqrt{1+3\a \bigl\lb (1-8\a M)\rho - 2M\bigr\rb}}\Bigr\rb 
\lab{sound-speed}
\ee
with $\rho$ as in \rf{rho-A}-\rf{cubic-eq}. Obviously $c_s^2 \to 0$ for $\a \to 0$, 
which conforms to the limiting values
\rf{rho-small-alpha}-\rf{p-small-alpha}. Also, $c_s^2$ \rf{sound-speed} is 
well-defined (positive) and reasonably small for $\rho > 2M (1-8\a M)^{-1}$, 
which is obviously fulfilled for small $\a$ according to \rf{rho-small-alpha}.

The new aspects in our current treatment w.r.t. 
Refs.\ct{scherrer}-\ct{pure-k-essence-5} are as follows:
\begin{itemize}
\item
We have derived an explicit duality between the original $g_{\m\n}$-frame 
quadratic $f(R)$-gravity plus non-Riemannian-modified-measure scalar field action 
\rf{TMT-A}, on one hand, and the Einstein-frame special quadratic purely kinetic 
``k-essence'' theory \rf{k-essence-A}.
\item
Unlike the general purely kinetic ``k-essence'' treatment 
\ct{scherrer}-\ct{pure-k-essence-5}
all three coefficients in our kinetic ``k-essence'' action 
\rf{L-eff}-\rf{simplifying-notat}
explicitly depend on only two independent parameters $(\a, M)$, the latter being a
dynamically generated integration constant.
\item
In spite of the singularity of all ``k-essence'' coefficients $A_1, A_2, A_3$ 
\rf{simplifying-notat}
for small $\a$ (the coupling constant of the $R^2$ in the original theory
\rf{TMT-A}), 
the energy density and pressure in the dual purely kinetic 
``k-essence'' theory have smooth limit for $\a \to 0$, whereby their limiting values
\rf{rho-small-alpha} and \rf{p-small-alpha}
exactly coincide with the corresponding values in the original theory
\rf{TMT-A} with $\a=0$, \textsl{i.e.}, \rf{TMT-0} \ct{dusty}.
\item
The established duality between \rf{TMT-A} and \rf{k-essence-A} elucidates the 
origin of the unified description of dark energy and dark matter within the
approach based on ``purely kinetic k-essence'' models \ct{scherrer}.
\end{itemize}

\section{Canonical Hamiltonian Formalism and Wheeler-DeWitt Equation}
\label{app}

For the canonical Hamiltonian formalism applied to generic generalized 
gravity-matter models with one or more non-Riemannian spacetime volume-forms
we refer to 
\ct{buggy}. In particular, for the theory \rf{TMT-0} 
the canonical Hamiltonian analysis was discussed in 
detail in Sec.2 of Ref.\ct{dusty}.

Here we will consider specifically the Hamiltonian treatment and quantization 
of the action \rf{FLRW-action-A} -- reduction of the purely kinetic ``k-essence'' 
model \rf{k-essence-A} for the Friedman-Lemaitre-Robertson-Walker 
(FLRW) class of metrics \rf{FLRW}.

From the explicit form of the FLRW action \rf{FLRW-action-A} we deduce the
canonically conjugated momenta $p_a$, $\pi_N$ and $p_\phi$ w.r.t. $a$, $N$ and 
$\phi \equiv {\wti \vp}$:
\br
p_a = - \frac{2a\adot}{N} \quad ,\quad \pi_N = 0 \; , \phantom{aaa}
\nonu \\
p_\phi = a^3 \Bigl\lb - \frac{1}{2\a} \frac{\phidot}{N} + 
\bigl(\frac{1}{4\a} - 2M\bigr) \frac{\phidot^3}{N^3} \Bigr\rb \; ,
\lab{can-mom}
\er
where the second relation for $\pi_N$ is a primary Dirac first-class constraint.
The total canonical Hamiltonian becomes:
\be
\cH_{\rm total} = N \Bigl\lb - \frac{p_a^2}{24a} + a^3 \rho (p_\phi/a^3)\Bigr\rb
\lab{can-Ham-reduced}
\ee
with $\rho (p_\phi/a^3)$ as in \rf{rho-A}, thus $\cH_{\rm total}$ by itself is 
a secondary Dirac first-class constraint with $N$ playing the role ot its 
Lagrange multiplier.
Since $\phi \equiv {\wti \vp}$ is a cyclic variable its canonically
conjugated momentum $p_\phi$ is conserved.

Quantization according to Dirac constrained Hamiltonian formalism
proceeds by imposing the quantized 
operator version of the Hamiltonian constraint -- the expression in the
square brackets in \rf{can-Ham-reduced} -- on the quantum wave function
$\Psi (a,p_\phi)$ -- the Wheeler-DeWitt equation. The ordering ambiguity in the
quantized version of the first term there is resolved by changing variables:
\be
a \to {\wti a} = \frac{4}{\sqrt{3}} a^{3/2} \; ,
\lab{a-new}
\ee
and taking the special operator ordering:
\be
\frac{p_a^2}{24 a} \to 
\frac{1}{\sqrt{12 a}}{\widehat p}_a \frac{1}{\sqrt{12 a}}{\widehat p}_a
= -\h \frac{\pa^2}{\pa {\wti a}^2} \; .
\lab{operator-order}
\ee
Therefore, the Wheeler-DeWitt equation acquires the form of Schr{\"o}dinger
equation for zero energy eigenvalue:
\be
\Bigl\lb - \h \frac{\pa^2}{\pa {\wti a}^2} + \cV_{\rm eff}({\wti a}, p_\phi)\Bigr\rb
\Psi ({\wti a}, p_\phi) = 0 
\lab{WDW-eq}
\ee
with effective potential:
\be
\cV_{\rm eff}({\wti a}, p_\phi) = - a^3 \rho (p_\phi/a^3)
\lab{eff-potential-0}
\ee
($\rho (p_\phi/a^3)$ as in \rf{rho-A}, and $a$ and ${\wti a}$ related as in
\rf{a-new}). Explicitly:
\be
\cV_{\rm eff}({\wti a}, p_\phi) = - \frac{3{\wti a}^2}{16} \Bigl\lb \frac{1}{8\a}
f^2({\wti a}, p_\phi) + \frac{4 p_\phi}{{\wti a}^2}\, f({\wti a}, p_\phi)
- \frac{1}{4\a}\Bigr\rb \; ,
\lab{eff-potential}
\ee
where $f({\wti a}, p_\phi)\equiv y$ is a root of the cubic equation
(cf. Eqs.\rf{phi-eq}, \rf{cubic-eq}-\rf{cubic-root}:
\be
y^3 - \frac{2}{1-8\a M}\, y - \frac{64\a p_\phi}{3(1-8\a M)} {\wti a}^{-2} = 0 \; .
\lab{cubic-eq-1}
\ee

Using the asymptotics \rf{rho-small-a}-\rf{rho-large-a} we obtain for the 
small ${\wti a}$ and large ${\wti a}$ of $\cV_{\rm eff}({\wti a}, p_\phi)$:
\be
\cV_{\rm eff}({\wti a}, p_\phi) \simeq 
- \Bigl(\frac{9 \a p_\phi^4}{1-8\a M}\Bigr)^{1/3} {\wti a}^{-2/3} 
- \Bigl(\frac{2\a p_\phi}{1-8\a M}\Bigr)^{2/3} \frac{1}{8\a}
\lab{small-wti-a}
\ee
for small ${\wti a}$, and:
\be
\cV_{\rm eff}({\wti a}, p_\phi) \simeq - \frac{3M}{8(1-8\a M)} {\wti a}^2
- \sqrt{\frac{2}{1-8\a M}}\, p_\phi 
\lab{large-wti-a}
\ee
for large ${\wti a}$. Using \rf{cubic-eq-1} we find that
$\cV_{\rm eff}({\wti a}, p_\phi)$ \rf{eff-potential} is negative for all 
${\wti a} \in (0,\infty)$ and diverges to $-\infty$ at both ends of the
interval according to \rf{small-wti-a}-\rf{large-wti-a}.




According to Scherrer's inequality \rf{scherrer} $\a$ must be very small.
Thus, substituting \rf{rho-small-alpha} into \rf{eff-potential-0} and
accounting for the change of variables \rf{a-new} brings the Wheeler-DeWitt
effective potential to the following form:
\br
\cV_{\rm eff}({\wti a}, p_\phi) \simeq - \frac{3M}{8}(1+8\a M)\,{\wti a}^2
\nonu \\
- \a \frac{8 p_\phi^2}{3}\,{\wti a}^{-2} - \sqrt{2} p_\phi (1+4\a M) \; ,
\lab{eff-potential-1}
\er
valid for any ${\wti a} \in (0,\infty)$.
In other words, for small $\a$ \rf{eff-potential-1} is a sum of {\em inverted}
harmonic oscillator potential with negative frequency squared:
\be
\om^2 = - \frac{3M}{4}(1+8\a M) = - \frac{3}{4} \L_{\rm eff} \; ,
\lab{freq-negative}
\ee
where $\L_{\rm eff}$ is the effective cosmological constant \rf{CC-eff}
for small $\a$, plus an inverse square potential and with an 
``energy'' eigenvalue $\cE = \sqrt{2} p_\phi (1+4\a M)$.

Thus, in the limit $\a \to 0$ the Wheeler-DeWitt equation \rf{WDW-eq} 
becomes Schr{\"o}dinger equation for an {\em inverted} harmonic oscillator
with negative frequency squared \rf{freq-negative} and 
with ``energy'' eigenvalue $\cE = \sqrt{2} p_\phi$:
\be
\Bigl\lb - \h \frac{\pa^2}{\pa {\wti a}^2} - \frac{3M}{8}\,{\wti a}^2 
- \sqrt{2} p_\phi \Bigr\rb \Psi ({\wti a}, p_\phi) = 0 \; .
\lab{WDW-eq-1}
\ee

The inverted harmonic oscillator was extensively studied in Ref.\ct{barton}
(for a more recent account and further references, see \ct{bermudez}). In
particular, the inverted oscillator was applied in \ct{guth-pi}
to study the quantum mechanical dynamics of the scalar field in the so
called ``new inflationary'' scenario. Since the energy eigenvalue spectrum 
of the inverted harmonic oscillator is continuous ($\cE \in (-\infty,+\infty)$)
and the corresponding energy eigenfunctions are not square-integrable, its
application in the context of cosmology \ct{guth-pi} required employment of
wave-packets w.r.t. $\cE$ instead of energy eigenfunctions. 

Similarly, in the present
case the value of $p_\phi$ -- the conserved canonical momentum of the kinetic
``k-essence'' scalar field $\phi \equiv {\wti \vp}$ in \rf{FLRW-action-A} --
is the analog of energy eigenvalue $\cE$ in the
Wheeler-DeWitt Schr{\"o}dinger-like equation \rf{WDW-eq-1} (modulo the factor
$\sqrt{2}$). Therefore, now in the strong coupling limit $\a \to 0$ 
the ``k-essence'' field $\phi \equiv {\wti \vp}$
plays the role of a Wheeler-DeWitt time $\t \equiv \frac{1}{\sqrt{2}}\phi$
and Eq.\rf{WDW-eq-1} acquires the form of a ``time-dependent''
Schr{\"o}dinger-like equation for the inverted harmonic oscillator upon
Fourier transforming the ``energy''-eigenvalue Eq.\rf{WDW-eq-1}:
\br
i\frac{d}{d\t} \Psi ({\wti a}, \t) = 
\Bigl\lb - \h \frac{\pa^2}{\pa {\wti a}^2} - \frac{3M}{8}\,{\wti a}^2\Bigr\rb
\Psi ({\wti a}, \t) \; ,
\lab{WDW-eq-2} \\
\Psi ({\wti a}, \t) = \int_{-\infty}^{\infty} d p_\phi \frac{1}{\sqrt{2}\pi}
e^{-i \sqrt{2} p_\phi \t} \Psi ({\wti a}, p_\phi) \; .
\lab{wave-packet}
\er
Following \ct{guth-pi} the appropriate normalized to unity (on the semiaxis 
${\wti a} \in (0,\infty)$) wave-packet solution of the ``time-dependent''
Wheeler-DeWitt equation \rf{WDW-eq-2} is of the form 
(using notation $\om$ \rf{freq-negative} for $\a=0$):
\br
\Psi ({\wti a}, \t) = \Bigl(\frac{2\om}{\pi} \sin(2b)\Bigr)^{1/4}
\bigl(\cos(b-i\om\t)\bigr)^{-1/2}
\nonu \\
\times \exp\{-\h {\wti a}^2 \om \tan(b-i\om\t)\} \; ,
\lab{WDW-sol}
\er
where $b$ is an integration constant describing the width of the wave packet.
Accordingly, the average value of the FLRW scale factor 
$a = \frac{\sqrt{3}}{4} {\wti a}^{2/3}$ (cf. \rf{a-new}-\rf{operator-order})
is given by:
\be
\llangle {\wti a} \rrangle \equiv \int_0^{\infty} d{\wti a}\, {\wti a}
|\Psi ({\wti a}, \t)|^2 
= \Bigl\lb \frac{\cos(2b) + \cosh(2\om \t)}{\pi \om \sin(2b)}\Bigr\rb^{1/2} \; ,
\lab{a-average}
\ee
exhibiting no singularity ($\llangle {\wti a} \rrangle \to 0$) at any ``time'' $\t$.

\section{Conclusions}
\label{conclude}

In the present paper we have discussed in some detail the main properties of
a generalized model of gravity interacting with a single scalar field, where
we have employed the method of non-Riemannian spacetime volume-forms
(alternative generally-covariant integration measure densities) constructed
in terms of auxiliary maximal rank tensor gauge fields (``measure'' gauge fields).

In the preceding paper \ct{dusty} we have shown that the 
non-Riemannian-measure-modified scalar field action \rf{TMT-0} yields a simple
unified description of dark energy and dust dark matter. Namely, the
corresponding energy density arises as an exact sum of a dark energy component
in the form of a dynamically generated cosmological constant appearing as an
arbitrary integration constant in the solution of the ``measure'' gauge
field equations, and a dark matter component produced by a hidden Noether symmetry
(not affecting the gravity part) giving rise to a Noether conserved current, 
which identifies the scalar field dynamics as a dust fluid motion along geodesics.

Here we extended the above treatment by coupling the 
non-Riemannian-measure-modified scalar field dynamics to quadratic $f(R)$
gravity. We have found an explicit duality in the usual sense of ``weak versus
strong coupling'' between the original non-standard 
gravity-scalar-field model providing exact unified description of dynamical 
dark energy and dust fluid dark matter in the matter sector, on one hand, 
and a quadratic purely kinetic ``k-essence'' gravity-matter model, on the other hand. 
The latter dual theory arises as the ``Einstein-frame'' theory of its original
counterpart. It is special in a sense that the couplings in the dual quadratic 
kinetic ``k-essence'' action are given in terms of only two parameters 
$(\a, M)$ -- the $R^2$-coupling constant in the original action \rf{TMT} and
a dynamically generated integration constant $M$ upon solving the equations
of motion for the auxiliary ``measure'' gauge field in \rf{TMT}. Moreover,
in spite of the divergence of the ``k-essence'' coupling constants  when 
$\a \to 0$ (strong coupling limit), both the ``k-essence''  energy density 
and ``k-essence'' pressure have smooth limits at $\a=0$ with the limiting values
coinciding with their respective values in \rf{TMT-0} -- the weak coupling
limit of the original theory \rf{TMT}, delivering an
explicit unified description of dark energy and dust dark matter as an exact sum of
two separate contributions to the total energy density.

The established duality in the present paper explains the ability of the
purely kinetic ``k-essence'' models \ct{scherrer} to provide {\em
approximately} a unified description of dark energy and dark matter and reveals 
that this unified description becomes {\em exact} in the strong coupling limit 
of a special type of quadratic purely kinetic ``k-essence'' theory. 

Finally, we have used the standard Dirac approach to constrained
Hamiltonian systems for a canonical Hamiltonian treatment of the 
non-Riemannian-measure-modified gravity-scalar-field theory, specifically
for the reduction of the latter in the case of FLRW class of cosmological 
spacetime metrics. In the limit of vanishing $R^2$ coupling 
the associated Wheeler-DeWitt equation acquires the form of a
Schr{\"o}\-din\-ger-like equation with an effective
potential of inverted harmonic oscillator. The quantum average value of the
FLRW scale factor does not exhibit any singularities in its time evolution.

\begin{acknowledgements}
We gratefully acknowledge support of our collaboration through the 
academic exchange agreement between the Ben-Gurion University in Beer-Sheva,
Israel, and the Bulgarian Academy of Sciences. 
S.P. and E.N. have received partial support from European COST actions
MP-1210 and MP-1405, respectively, as well from Bulgarian National Science
Fund Grant DFNI-T02/6. 
We also thank the referees for useful remarks.
\end{acknowledgements}


\end{document}